\documentclass{article}

\usepackage{graphics,epsfig,here}

\def\greaterthansquiggle{\raise.3ex\hbox{$>$\kern-.75em\lower1ex\hbox{$\sim$}}}
\def\lessthansquiggle{\raise.3ex\hbox{$<$\kern-.75em\lower1ex\hbox{$\sim$}}}
\newcommand{\beq}{\begin{equation}}
\newcommand{\eeq}{\end{equation}}
\newcommand{\beqa}{\begin{eqnarray}}
\newcommand{\eeqa}{\end{eqnarray}}
\newcommand{\ba}{\begin{array}}
\newcommand{\ea}{\end{array}}

\def\ti    {\tilde}

\def\st    {{\ti t}}

\newcommand{\CH}{\tilde\chi}

\newcommand{\kst}{k^{\st}}
\newcommand{\lst}{l^{\st}}
\newcommand{\ksn}{k^{\SNT}}
\newcommand{\lsn}{l^{\SNT}}
\newcommand{\SNT}{\ti\nu_{\tau}}
\newcommand{\T}{\ti\tau}

\newcommand{\gsim}{\;\raisebox{-0.9ex}
           {$\textstyle\stackrel{\textstyle >}{\sim}$}\;}
\newcommand{\lsim}{\;\raisebox{-0.9ex}{$\textstyle\stackrel{\textstyle <}
           {\sim}$}\;}

\begin{document}  
\setlength{\unitlength}{1mm}

\begin{flushright}
  UWThPh-2001-36\\
  HEPHY-PUB 746 \\
  ZH-TH 5/02\\
  hep-ph/0202198\\[3mm]
\end{flushright}

\begin{center}

{\Large \bf\boldmath A CP sensitive asymmetry in the three--body decay $\tilde t_1\to b \SNT \tau^+$}

\vspace{5mm}

{\large A.~Bartl,$^1$~ 
T.~Kernreiter,$^1$~
W.~Porod\,$^{3,4}$} \\

\vspace{4mm}

{\normalsize \it
$^1$~Institut f\"ur Theoretische Physik, Universit\"at Wien, 
     A--1090 Vienna, Austria \\
$^3$~Inst.~f.~Hochenergiephysik, \"Oster.~Akademie d.~Wissenschaften,
      A-1050 Vienna, Austria \\
$^4$~Inst.~f\"ur Theor. Physik, Universit\"at Z\"urich, CH-8057 Z\"urich,
      Switzerland}

\end{center}

\begin{abstract} 
We consider the three--body decay $\tilde t_1\to b \SNT \tau^+$ and
propose the asymmetry of the $\tau$ polarization perpendicular
to the decay plane as a $CP$ sensitive observable. We calculate
this asymmetry in the Minimal Supersymmetric Standard Model with
the parameters $\mu$ and $A_t$ complex. In the parameter domain where
the decay $\tilde t_1\to b \SNT \tau^+$ is important this asymmetry
can go up to $\pm 30\%$. We also estimate the event rates 
necessary to observe this asymmetry at $90\%$ CL.
\end{abstract}

\section{Introduction}
The experimental search for supersymmetric (SUSY) particles will have
high priority at the upgraded Tevatron and at LHC. The precision
determination of the SUSY parameters will be the main goal of a future
$e^+e^-$ linear collider \cite{acco}.  The analysis of scalar top
quarks $\st_i, i=1,2$, will be particularly interesting, because of
the large top Yukawa coupling involved in this system. Due to the
effects of the top Yukawa coupling the lighter stop may be relatively
light if not the lightest charged SUSY particle \cite{elru}.

A phenomenological study of production and decays of 3rd generation
sfermions at an $e^+e^-$ linear collider with cms energy in the range
$\sqrt{s}=0.5-1$~TeV has been given in Ref.\,\cite{sferm}. There it
has been shown that by measuring production cross sections with
polarized beams the masses of the top squarks $m_{\st_1}, m_{\st_2}$
and their mixing angle $\theta_{\st}$ can be determined. It has also
been shown that with an integrated luminosity of $500 fb^{-1}$ a
precision of about $1\%$ can be achieved.  The precision to be
expected for the underlying SUSY parameters has also been estimated.
This analysis has been performed in the Minimal Supersymmetric
Standard Model (MSSM) with real parameters.

However, the assumption that all MSSM parameters are real may be too
restrictive.  In principle, the Higgs--higgsino mass parameter $\mu$
and the trilinear scalar coupling parameters $A_f$ of the sfermions
$\ti f$ may be complex. These complex parameters are new sources of
$CP$ violation and may provide potentially large SUSY contributions to
the electric dipole moments (EDM) of electron and neutron.  The very
small experimental upper limits of the electron and neutron EDMs,
therefore, may lead to restrictions on the complex phases.  Recent
analyses have shown that in mSUGRA--type models the phase of $\mu$ is
restricted to $|\varphi_{\mu}|\lsim 0.1-0.2$ for a universal scalar
mass parameter $M_0\lsim$ 400~GeV, whereas the phase of the universal
trilinear scalar coupling parameter $A_0$ is correlated with
$\varphi_{\mu}$, but otherwise unrestricted \cite{edm}.  One can
conclude that in models with more general parameter specifications the
phases of the parameters $A_f$ of the 3rd generation sfermions are not
restricted at one--loop level by the electron and neutron EDMs.  
There may be restrictions at two--loop level \cite{pilaf}.
Furthermore, a complex trilinear coupling parameter $A_t$ in the stop
system can also lead to interesting $CP$ violating effects in top
quark production, as discussed in \cite{top}.

For a complete analysis of the stop sector one has to take into account
that the parameters $\mu$ and $A_t$ may be complex. The parameter
$|\mu|$ and its phase $\varphi_{\mu}$ will presumably be determined by measuring
suitable observables of the chargino and neutralino sector \cite{kali}. For
the determination of $|A_t|$ and its phase $\varphi_{A_t}$ appropriate
observables in the stop sector have to be measured. However, it may be difficult 
to define a suitable
$CP$ sensitive observable in stop decays if the main decay
modes are two--body decays.

In the present paper we define a $CP$ sensitive asymmetry in the decay 
$\tilde t_1\to b \SNT  \tau^+$. As this is a three--body decay the polarization
of the $\tau^+$ normal to the decay plane is sensitive to $CP$ violation.
The appropriate $CP$ sensitive observable is defined by the
asymmetry of the $\tau$ polarization perpendicular to the decay plane.
As we will show this asymmetry can go up to $30\%$. 
Moreover, we show the existence of parameter regions where the decay
$\tilde t_1\to b \SNT  \tau^+$ has a sufficient branching ratio allowing for
the measurement of this asymmetry.
We perform our analysis in the MSSM with $\mu$ and $A_t$ complex.
We focus on scenarios where only the decays 
$\tilde t_1\to b \ti\nu_{\ell} \ell,\
\tilde t_1\to b \ti\ell \nu_{\ell}$ \cite{werner}, 
$\tilde t_1\to c \CH_1^0$ \cite{hiko} and four body decays \cite{djo}
are kinematically allowed.
We assume that 
the lightest neutralino $\CH^0_1$ is the lightest SUSY particle (LSP). 
The $CP$ asymmetry defined above is analogous to that
defined in Ref.\,\cite{atwood} in case of top quark decays.

In Section~2 we shortly review stop mixing in the presence of complex
parameters. 
In Section~3 we give the formulae of the
CP violating observable.
In Section~4 we present numerical results for the phase dependences of the CP asymmetry.
We give an theoretical estimate of the event rates  necessary to observe the 
$CP$ sensitive asymmetry at an $e^+e^-$ linear collider with $\sqrt{s}=$ 
0.5 -- 1~TeV.
Section~5 contains a short summary.

\section{$\tilde t_L$ -- $\tilde t_R$ Mixing}

We first give a short account of $\tilde t_L$ -- $\tilde t_R$ mixing in the
case the parameters $\mu$ and
$A_t$ are complex. The masses and couplings of the
$\st$--squarks follow from the 
hermitian $2 \times 2$ mass matrix which in the basis 
$(\st_L, \st_R)$ reads
\begin{equation}
{\mathcal{L}}_M^{\st}= -(\st_L^{\ast},\, \st_R^{\ast})
\left(\begin{array}{ccc}
M_{\st_{LL}}^2 & e^{-i\varphi_{\st}}|M_{\st_{LR}}^2|\\[5mm]
e^{i\varphi_{\st}}|M_{\st_{LR}}^2| & M_{\st_{RR}}^2
\end{array}\right)\left(
\begin{array}{ccc}
\st_L\\[5mm]
\st_R \end{array}\right),
\label{eq:mm}
\end{equation}
where
\begin{eqnarray}
M_{\st_{LL}}^2 & = & M_{\tilde Q}^2+(\frac{1}{2}-\frac{2}{3}\sin^2\Theta_W)
\cos2\beta \ m_Z^2+m_t^2 ,\\[3mm]
M_{\st_{RR}}^2 & = & M_{\tilde U}^2+\frac{2}{3}\sin^2\Theta_W\cos2\beta \
m_Z^2+m_t^2 ,\\[3mm]
M_{\st_{RL}}^2 & = & (M_{\st_{LR}}^2)^{\ast}=
m_t(A_t-\mu^{\ast}  
\cot\beta), \label{eq:mlr}
\end{eqnarray}
\begin{equation}
\varphi_{\st}  = \arg\lbrack A_\st-\mu^{\ast}\cot\beta\rbrack ,
\label{eq:phtau}
\end{equation}
where $\tan\beta=v_2/v_1$ with $v_1 (v_2)$ being the vacuum 
expectation value of the Higgs field $H_1^0 (H_2^0)$,
$m_t$ is the mass of the top quark and 
$\Theta_W$ is the weak mixing angle, $\mu$ is the Higgs--higgsino mass parameter
and $M_{\ti Q}$, 
$M_{\ti U}, A_t$ are the soft SUSY--breaking parameters of the stop
system. 
The mass eigenstates $\st_i$ are $(\ti t_1, \ti t_2)=
(\st_L, \st_R) {\mathcal{R}^{\st}}^T$ with
 \begin{equation}
\mathcal{R}^{\st}=\left( \begin{array}{ccc}
e^{i\varphi_{\st}}\cos\theta_{\st} & 
\sin\theta_{\st}\\[5mm]
-\sin\theta_{\st} & 
e^{-i\varphi_{\st}}\cos\theta_{\st}
\end{array}\right),
\label{eq:rtau}
\end{equation}
with
\begin{equation}
\cos\theta_{\st}=\frac{-|M_{\st_{LR}}^2|}{\sqrt{|M_{\st _{LR}}^2|^2+
(m_{\st_1}^2-M_{\st_{LL}}^2)^2}},\quad
\sin\theta_{\st}=\frac{M_{\st_{LL}}^2-m_{\st_1}^2}
{\sqrt{|M_{\st_{LR}}^2|^2+(m_{\st_1}^2-M_{\st_{LL}}^2)^2}}.
\label{eq:thtau}
\end{equation}
\\
The mass eigenvalues are
\begin{equation}
 m_{\st_{1,2}}^2 = \frac{1}{2}\left((M_{\st_{LL}}^2+M_{\st_{RR}}^2)\mp 
\sqrt{(M_{\st_{LL}}^2 - M_{\st_{RR}}^2)^2 +4|M_{\st_{LR}}^2|^2}\right).
\label{eq:m12}
\end{equation}
\\
$\tilde t_L$ -- $\tilde t_R$ mixing is naturally large because of the large 
top quark mass entering in the off--diagonal elements 
of the mass matrix
(see Eqs.\,(\ref{eq:mm}) and (\ref{eq:mlr})). This is important for the $CP$ sensitive
observable discussed below, 
because it is proportional to $\sin\theta_{\st} \cos\theta_{\st}$.
Note further that
for $|A_t|\gg |\mu|\cot\beta$ we have $\varphi_{\st} \approx
\varphi_{A_t}$. 
%
\section{Tau Polarization Asymmetry}
%
The parts of the Lagrangian relevant for the three--body 
decay $\tilde t_1\to b \SNT  \tau^+$ are
\begin{equation}
\mathcal{L}_{b\CH_j\st_1}=g \bar b(\kst_{1j}P_L+\lst_{1j}P_R)\CH_j^{c+}\st_1+h.c.,
\end{equation}
\begin{equation}
\mathcal{L}_{\tau\CH_j\SNT}=g \overline{\CH_j^{c+}} ({\ksn_j}^{\ast}P_R+
{\lsn_j}^{\ast}P_L)\tau\ \SNT^{\ast}+h.c.,
\end{equation}
where
\begin{equation}
\label{coupst}
\lst_{1j}= -e^{-i\varphi_{\st}}\cos\theta_{\st} V_{j1}+ 
 Y_t \sin\theta_{\st} V_{j2},\qquad
\kst_{1j}=Y_b e^{-i\varphi_{\st}}\cos\theta_{\st} U_{j2}^{\ast},
\end{equation}
\begin{equation}
\lsn_{j}=-V_{j1},\qquad
\ksn_{j}=Y_{\tau} U_{j2}^{\ast}.
\label{coupsn}
\end{equation}
$g$ is the 
weak coupling constant, $P_{R,L}=1/2(1\pm\gamma_5)$ and 
the Yukawa
couplings are $Y_t=m_t/\sqrt{2}m_W\sin\beta$, $Y_b=m_b/\sqrt{2}m_W\cos\beta$,
$Y_{\tau}= m_{\tau}/\sqrt{2}m_W\cos\beta$.
The unitary $2\times 2$ matrices $U$ and $V$ diagonalize the chargino
mass matrix \cite{haka}. A convenient parametrization of $U$ and $V$ for 
complex parameters
can be found in \cite{oshimo}. 

The three body decay $\st_1\to b \SNT \tau^+$ proceeds via exchange of
charginos
$\CH_i^{\pm}, i=1,2$. 
We consider the
polarization of the outgoing $\tau^+$ perpendicular to the decay plane,
which is sensitive to $CP$ violation.
We define the unit vector 
\begin{equation}
\vec{e}_N=\frac{\vec{p}_{\tau}\times\vec{p}_b}{|\vec{p}_{\tau}\times\vec{p}_b|}.
\label{unit}
\end{equation}
The average polarization of the outgoing $\tau^+$ in the direction $\vec{e}_N$ 
in the decay
$\tilde t_1\to b \SNT\tau^+$
is given by
\begin{equation}
\mathcal{P}_N^{\tau^+}=\frac{B(\tilde t_1\to b \SNT \tau^+(\vec{e}_N))-
B(\tilde t_1\to b \SNT \tau^+(-\vec{e}_N))}
{B(\tilde t_1\to b \SNT \tau^+(\vec{e}_N))+
B(\tilde t_1\to b \SNT \tau^+(-\vec{e}_N))}.
\label{polantiL}
\end{equation}
For the CP conjugated process we get
\begin{equation}
\mathcal{P}_N^{\tau^-}=\frac{B(\bar{\st}_1\to \bar b \bar{\ti \nu}_{\tau}
\tau^-(\vec{e}_N))-
B(\bar{\st}_1\to \bar b \bar{\ti \nu}_{\tau}
\tau^-(-\vec{e}_N))}
{B(\bar{\st}_1\to \bar b \bar{\ti \nu}_{\tau}
\tau^-(\vec{e}_N))+
B(\bar{\st}_1 \to \bar b \bar{\ti \nu}_{\tau}
\tau^-(-\vec{e}_N))}.
\label{polL}
\end{equation}
Note that $\mathcal{P}_N^{\tau^-}=-\mathcal{P}_N^{\tau^+}$ since the couplings
are complex conjugate to each other.
The observables $\mathcal{P}_N^{\tau^+}$ and $\mathcal{P}_N^{\tau^-}$ 
are odd under naive 
time reversal $T_{\mathcal{N}}$, where only the polarization and momentum
vectors are reversed but initial 
and final states are not interchanged. We can define a $CP$ sensitive
asymmetry of the form
\begin{equation}
\mathcal{A}_{CP}=\frac{1}{2}(\mathcal{P}_N^{\tau^+}-\mathcal{P}_N^{\tau^-}).
\label{CPsensi}
\end{equation}
In order to obtain a $CP$ asymmetry which for practical reasons
is more useful, we have to take into account the subsequent decay 
$\SNT\to \CH_1^0 \nu_{\tau}$. The complete decay chain is then
$\tilde t_1\to b \SNT \tau^+ \to b \tau^+ \CH_1^0 \nu_{\tau}$.
Another decay chain leading to the same final state is 
$\tilde t_1\to b \T_i \nu_{\tau}\to b \tau \CH^0_1 \nu_{\tau}$,
where in the second step the $\T_i$ decays into $\tau \CH_1^0$.
In the $\ti\tau_i$ rest system $\ti\tau_i\to\tau\CH^0_1$ has an isotropic
decay distribution.
The decay mode $\tilde t_1\to b \T_i \nu_{\tau}\to b \tau \CH^0_1 \nu_{\tau}$
can be easily incorporated in our
consideration.
 
However, the decay $\tilde t_1\to b W \CH_1^0\to b \tau \CH^0_1 \nu_{\tau}$
also leads to the same final state~\cite{powo}.
The decay $\tilde t_1\to b W \CH_1^0\to b \tau \CH^0_1 \nu_{\tau}$
is more involved,
because the $W$ polarization leads to a non-vanishing correlation 
between the $\vec{p}_{\tau}$ -- $\vec{p}_b$ plane and the $\tau$ polarization. 
While this may also lead to $CP$ sensitive effects, in the present
paper we confine ourselves to the discussion of the $CP$ asymmetry in the decay
chains $\tilde t_1\to b \SNT \tau \to b \tau \CH_1^0 \nu_{\tau}$
and $\tilde t_1\to b \T_i \nu_{\tau}\to b \tau \CH^0_1 \nu_{\tau}$, assuming
that $m_{\st_1}<m_W+m_{\CH_1^0}+m_b$.

We define a $CP$ asymmetry  similar to Eq.\,(\ref{CPsensi}), but for
the final state $b \tau \CH^0_1 \nu_{\tau}$:
\begin{equation}
\mathcal{A}^{\prime}_{CP} = \frac{1}{2}
(\mathcal{P^{\prime}}_N^{\tau^+}-\mathcal{P^{\prime}}_N^{\tau^-}),
\label{Asy}
\end{equation}
with
\begin{equation}
\mathcal{P^{\prime}}_N^{\tau^+} = \frac{
B(\tilde t_1\to f(\vec{e}_N))-B(\tilde t_1\to f(-\vec{e}_N))}
{B(\tilde t_1\to f(\vec{e}_N))+B(\tilde t_1\to f(-\vec{e}_N))},\quad
\mathcal{P^{\prime}}_N^{\tau^-} = \frac{
B(\bar{\st}_1\to \bar{f}(\vec{e}_N))-B(\bar{\st}_1\to \bar{f}(-\vec{e}_N))}
{B(\bar{\st}_1\to \bar{f}(\vec{e}_N))+B(\bar{\st}_1\to \bar{f}(-\vec{e}_N))},
\label{Asyzusatz}
\end{equation}
where we have introduced a shorthand notation $f(\vec{e}_N)\equiv 
\CH_1^0\bar\nu_{\tau} b \ \tau^+(\vec{e}_N)$ and
$\bar{f}(\vec{e}_N)\equiv 
\CH_1^0\nu_{\tau} \bar b \ \tau^-(\vec{e}_N)$. 
The branching ratio for the decay
$\st_1\to\CH_1^0\nu_{\tau} b \ \tau$ is to a good approximation given by
\begin{equation}
B(\tilde t_1\to \CH_1^0\nu_{\tau} b \ \tau)\simeq B(\tilde t_1\to b \SNT \tau)
B(\SNT\to\CH_1^0\nu_{\tau})+
\sum^2_{i=1} B(\tilde t_1\to b \T_i \nu_{\tau})B(\T_i\to\CH_1^0\tau).
\label{Br}
\end{equation}
This can also be seen by using the formulas for the 4--body stop decays 
given in~\cite{djo}.
Relation~(\ref{Br}) holds if $|m_{\T_i}-m_{\SNT}|\gg \Gamma_{\T_i}+\Gamma_{\SNT}$,
which is naturally fulfilled. If $B(\SNT\to\CH_1^0\nu_{\tau})\simeq 1$
and $B(\T_i\to\CH_1^0\tau^+)\simeq 1$, then Eq.(\ref{Asy}) can be rewritten as
\begin{equation}
\label{CPasy0}
\mathcal{A}^{\prime}_{CP} \simeq \frac{
\Gamma(\tilde t_1\to b \SNT \tau^+(\vec{e}_N))-
\Gamma(\tilde t_1\to b \SNT \tau^+(-\vec{e}_N))-
\Gamma(\bar{\st}_1\to \bar  b \bar{\ti \nu}_{\tau} \tau^-(\vec{e}_N))+
\Gamma(\bar{\st}_1\to 
\bar  b \bar{\ti \nu}_{\tau} \tau^-(-\vec{e}_N))}{2\ \Gamma_{\rm unpol}},
\end{equation}
where $\Gamma_{\rm unpol}=\Gamma(\tilde t_1\to b \SNT \tau)+
\sum^2_{i=1}\Gamma(\tilde t_1\to b \T_i \nu_{\tau})$.
A very useful approximation for $\mathcal{A}^{\prime}_{CP}$
can be obtained in the limit $\varphi_{\mu}\to 0$
\begin{equation}
\mathcal{A}^{\prime}_{CP} \simeq \frac{g^4}{\Gamma_{\rm unpol}}\ 2\ m_{\st_1} Y_t
 Y_{\tau}  |\sin2\theta_{\st}|\ |\mu|\ \sin\varphi_{\st}\ \mathcal{I},
\label{CPasy}
\end{equation}
where
\begin{equation}
\mathcal{I}= \frac{1}{2 m_{\st_1}}\int  \frac{d^3p_b}{(2 \pi)^3 2 E_b}  \frac{d^3p_{\tau}}
{(2 \pi)^3 2 E_{\tau}}
 \frac{d^3p_{\SNT}}{(2 \pi)^3 2 E_{\SNT}} (2\pi)^4\delta(p_{\st_1}-p_b-p_{\tau}-p_{\SNT}) 
 \frac{|\vec{p}_{\tau}||\vec{p}_b|
\sin\theta_{b\tau}}
{(p_{{\CH}^{\pm}}^2-m^2_{{\CH_1}^{\pm}})(p_{{\CH}^{\pm}}^2-m^2_{{\CH_2}^{\pm}})},
\end{equation}
where
$p_{\st_1}=(m_{\st_1},\vec{0}), p_{{\CH}^{\pm}}=p_{\st_1}-p_b$, $m_{{\CH_{1,2}}^{\pm}}$ 
are the masses of the charginos
and $\sin\theta_{b\tau}$ is the angle between the $b$ quark and the $\tau$ lepton.
Eq.\,(\ref{CPasy}) is very instructive, because it exhibits the main features of the
behaviour of $\mathcal{A}^{\prime}_{CP}$.
For example, the sign of $\mathcal{A}_{CP}^{\prime}$ is given by 
$sgn(\mathcal{A}_{CP}^{\prime})=sgn(\varphi_{\st})$.
The unpolarized decay width $\Gamma_{\rm unpol}$ appearing in Eq.\,(\ref{CPasy0}) and
(\ref{CPasy}) is given in \cite{werner} 
for real parameters and in \cite{kern} for
complex parameters.

Finally, we want to remark that
the absorptive part of the amplitude
caused by the Breit--Wigner form of the chargino propagators is strongly suppressed,
because we have assumed
that the decay $\st_1\to \CH^+_1b$ is not accessible. 
Therefore the rate asymmetry is expected to be at most of the order 
of $\approx 10^{-3}$.

\section{Numerical Results}
%
%
In the following we present numerical results for the $CP$ sensitive
asymmetry $\mathcal{A}_{CP}^{\prime}$ defined in Eqs.\,(\ref{Asy}) and
(\ref{Asyzusatz}).  Our input parameters are $m_{\st_1}, m_{\st_2},
m_{\SNT}, |A_t|, \varphi_{A_t}, A_\tau, \tan\beta, M_2, |\mu|$, assuming
$M_1=5/3\tan^2\Theta_W\ M_2$, with $M_1$ and $M_2$ real.  For
simplicity we set $\varphi_{\mu}=0$.  We impose the approximate
necessary condition for the tree--level vacuum stability
$|A_t|^2<3(M_{\ti Q}^2+M_{\ti U}^2+(m_A^2+m_Z^2)\cos^2\beta- m_Z^2/2)$
\cite{casas}.  For the pseudoscalar Higgs mass which appears in this
condition  we choose for definiteness $m_A = 150~$GeV.
We have checked that in the numerical examples studied below the restrictions
from the electron and neutron EDMs at two--loop level \cite{pilaf} are
fulfilled.

\begin{figure}[H]
\setlength{\unitlength}{1mm}
\begin{center}
\begin{picture}(150,60)
\put(-5,0){\mbox{\epsfig{figure=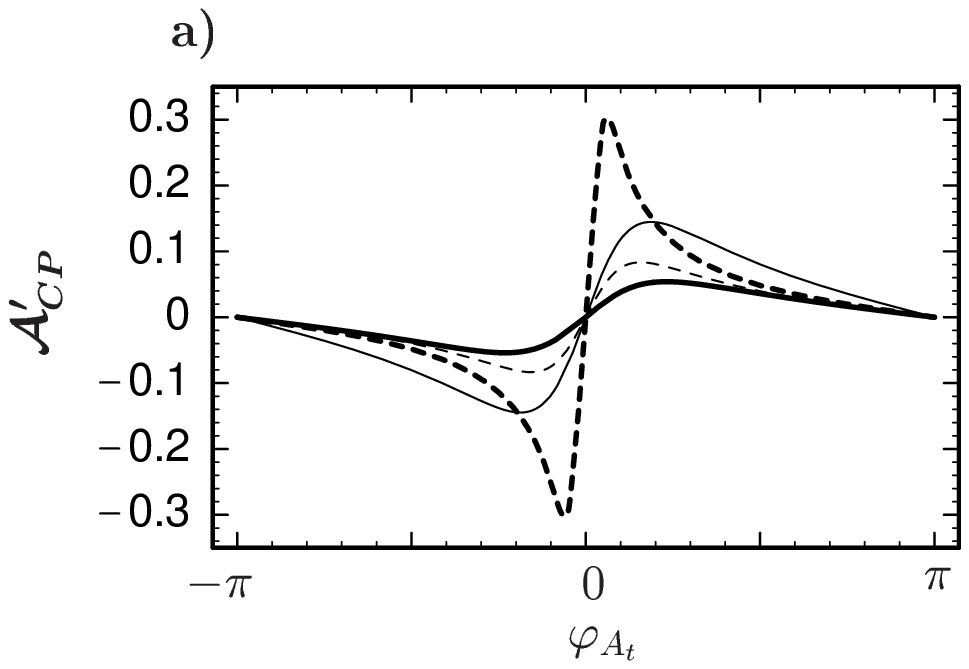,height=6.5cm,width=8.cm}}}
\put(75,0){\mbox{\epsfig{figure=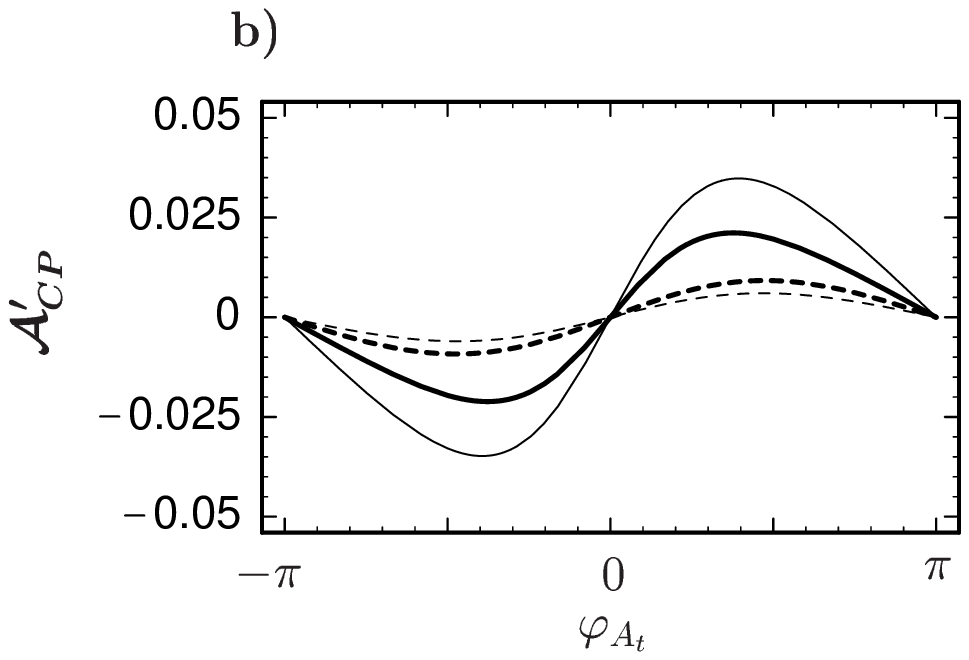,height=6.5cm,width=8.cm}}}
\end{picture}
\end{center}
\caption{The $CP$ sensitive asymmetry $\mathcal{A}^{\prime}_{CP}$ 
as a function of $\varphi_{A_t}$.
The input parameters are $m_{\st_1}= 240$~GeV, $m_{\st_2}=800$~GeV, $m_{\SNT}=200$~GeV,
$M_2=350$~GeV, $|A_t|=1000$~GeV,
$(\tan\beta=3,|\mu|=400$~GeV;~thick~solid~line),
$(\tan\beta=10,|\mu|=400$~GeV;~thin~solid~line),
$(\tan\beta=3,|\mu|=700$~GeV;~thick~dashed~line),
$(\tan\beta=10,|\mu|=700$~GeV;~thin~dashed~line),
for the cases: a)$ M_{\ti Q}>M_{\ti U}$
and b)$ M_{\ti Q}<M_{\ti U}$; }
\label{fig:cp1}
\end{figure}

\begin{figure}[H]
\begin{center}
\includegraphics{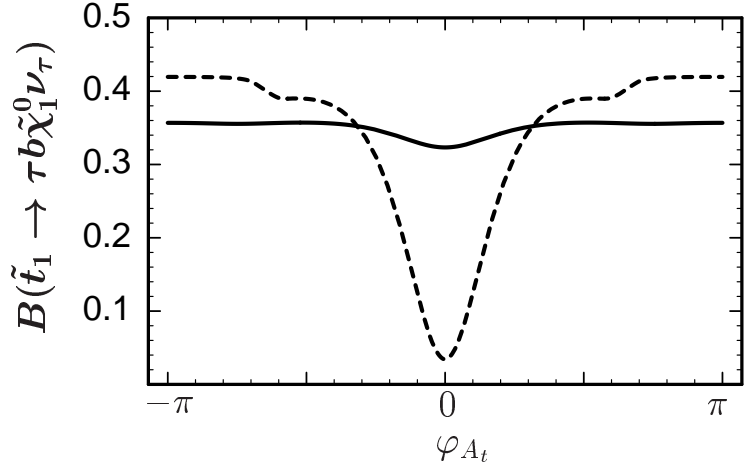}
\end{center}
\caption{The branching ratio of $\st_1\to\tau b\CH_1^0\nu_{\tau}$
as a function of $\varphi_{A_t}$, with 
$M_{\ti Q}>M_{\ti U}, m_{\st_1}=240$~GeV, $m_{\st_2}=800$~GeV, $m_{\SNT}=200$~GeV,
$\tan\beta=3,
M_2=350$~GeV, $|A_t|=1000$~GeV for
the two cases $|\mu|=$ 400~GeV~(solid~line), $|\mu|=$ 700~GeV~(dashed~line).}
\label{fig:cp2}
\end{figure}

First we consider the influence of the parameter $\varphi_{A_t}$.
In Figs.\,\ref{fig:cp1}a and \ref{fig:cp1}b we show 
$\mathcal{A}_{CP}^{\prime}$ as
a function of $\varphi_{A_t}$ for $M_{\ti Q}>M_{\ti U}
\ (|\cos\theta_{\st}|<|\sin\theta_{\st}|)$ and
$M_{\ti Q}<M_{\ti U} \ (|\cos\theta_{\st}|>|\sin\theta_{\st}|)$, respectively.
We display the asymmetry for the four scenarios
$(|\mu|=400$~GeV, $\tan\beta=3), (|\mu|=400$~GeV, $\tan\beta=10),
(|\mu|=700$~GeV, $\tan\beta=3)$ and $(|\mu|=700$~GeV, $\tan\beta=10)$,
taking $m_{\st_1}= 240$~GeV, $m_{\st_2}=800$~GeV, $m_{\SNT}=200$~GeV,
$M_2=350$~GeV.
We fix the mass of $\ti\tau_1$ by taking $M_{\ti E}=0.9 M_{\ti L}$.
As can be seen, the $CP$ asymmetry $\mathcal{A}_{CP}^{\prime}$ 
is much larger for $M_{\ti Q}>M_{\ti U}$
than for $M_{\ti Q}<M_{\ti U}$.
The reason is, that for $M_{\ti Q}<M_{\ti U}$ the $\st_1$ has a larger
$\st_L$ component than for $M_{\ti Q}>M_{\ti U}$, implying a stronger
coupling to the gaugino component of the charginos. This in turn implies a
larger $\Gamma_{\rm unpol}$ for $M_{\ti Q}<M_{\ti U}$.
Cearly $\mathcal{A}_{CP}^{\prime}$ has to vanish for the $CP$ conserving case
$\varphi_{A_t}= 0, \pm \pi$ (see the corresponding factor in Eq.(\ref{CPasy})).
For better understanding of the value of $\mathcal{A}_{CP}^{\prime}$ in the
region $|\varphi_{A_t}|\lsim \pi/4$, we show in Fig.\,\ref{fig:cp2} the branching ratio 
$B(\tilde t_1\to \CH_1^0\nu_{\tau} b \tau)$ as a function of $\varphi_{A_t}$, 
for the two scenarios $(|\mu|=400$~GeV, $\tan\beta=3)$ and $(|\mu|=700$~GeV, $\tan\beta=3)$,
keeping the other parameters as in Fig.\,\ref{fig:cp1}a.
The decay width $\Gamma(\tilde t_1\to \CH_1^0\nu_{\tau} b \tau)$ 
has a similar behaviour. The minimum at $\varphi_{A_t}=0$
can be traced back to the fact that $|\lst_{11}|$ (see Eq.\,(\ref{coupst})) 
has a minimum there, due to a negative interference between the gaugino
and higgsino contribution.

\begin{figure}[H]
\begin{center}
\includegraphics{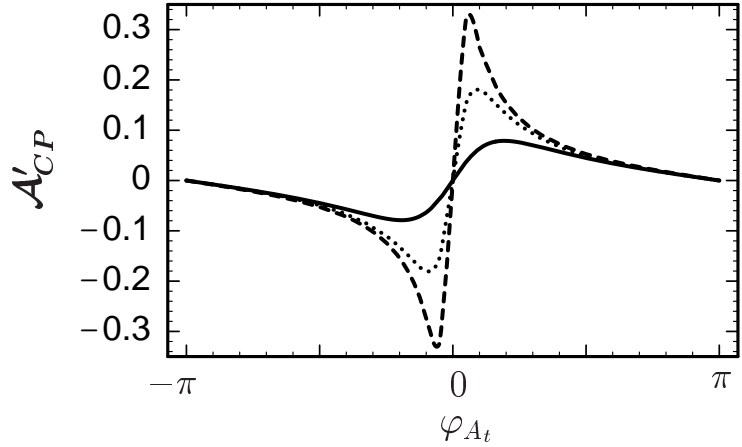}
\end{center}
\caption{The $CP$ sensitive asymmetry $\mathcal{A}^{\prime}_{CP}$ 
as a function of $\varphi_{A_t}$.
The input parameters are $m_{\st_1}= 240$~GeV, $m_{\st_2}=800$~GeV,$ m_{\SNT}=200$~GeV,
$M_2=350$~GeV, $\tan\beta=3,|\mu|=600$~GeV for $|A_t|=$ 600~GeV~(solid~line),
$|A_t|=$ 1000~GeV~(dashed~line), $|A_t|=$ 1300~GeV~(dotted~line). }
\label{fig:cp3}
\end{figure}

In Fig.\,\ref{fig:cp3} we show $\mathcal{A}_{CP}^{\prime}$ for three values 
of $|A_t|=$ 600~GeV, 1000~GeV and 1300~GeV, assuming $M_{\ti Q}>M_{\ti U}$.
The other parameters are
$\ m_{\st_1}= 240$~GeV, $m_{\st_2}=800$~GeV, $m_{\SNT}=200$~GeV,
$A_\tau = 0$,
$M_2=350$~GeV, $|\mu|=600$~GeV and $\tan\beta=3$.
For $|A_t|\lsim 1000$~GeV  $\mathcal{A}_{CP}^{\prime}$ increases with increasing
$|A_t|$, because $|\sin2\theta_{\st}|$ increases (see Eq.\,(\ref{CPasy})).
The decrease of $\mathcal{A}_{CP}^{\prime}$ for $|A_t|\gsim 1000$~GeV
is explained by the fact that $\Gamma_{\rm unpol}$ increases stronger
than $|\sin2\theta_{\st}|$.

The polarization of the $\tau$ is analysed through its decay distributions. 
Usually the decay modes \linebreak[4]
$\tau\to \pi\nu, \rho\nu, a_1\nu, \mu\nu\bar\nu, e\nu\bar\nu$ are
used as analyzers. As we are interested in the transverse polarisation
of the $\tau$ lepton, we  take only the $\rho\nu$ and $a_1\nu$ final states
for our analysis.
The sum of the branching ratios of these two decay modes is $\approx 34$\%.
We take for the sensitivities for measuring the polarization of the $\tau$ 
lepton the values quoted in Ref.\,\cite{davier} for an ideal detector.
Moreover, the numbers quoted are for longitudinal tau polarization and
it is expected that the sensitivities for transversely polarized tau leptons
are somewhat smaller. To account for both effects  
we assume a reduction of the sensitivity $S$ of
$10\%$ \cite{davierdiss}.
Furthermore, we assume that the direction of flight of the $\tau$ 
lepton can be reconstructed. Following \cite{davier}, the error 
in measuring the polarization asymmetry is given by 
\begin{equation}
\delta \mathcal{P^{\prime}}^{\tau^-}_N=\frac{1}{S_{\rm red}
\sqrt{N_{\tau}}},
\end{equation}
where $S_{\rm  red}= 0.3, N_{\tau}= B(\tilde t_1\to \CH_1^0\nu_{\tau} b  \tau) 
N_{\st_1}$, with $N_{\st_1}$ being the number of produced $\st_1 \bar{\st}_1$ pairs in the 
process $e^+ e^-\to \st_1\bar{\st}_1$. 

The minimum number $N_{\st_1}$ of produced $\st_1$ pairs necessary to observe 
$\mathcal{A}_{CP}^{\prime}$ at $90\%$ confidence level (CL) is then
given by
\begin{equation}
N_{\st_1}=\frac{1}{B(\tilde t_1\to \CH_1^0\nu_{\tau} b  \tau)}
\biggl(\frac{1.64}{S_{\rm red}|\mathcal{A}_{CP}^{\prime}|}\biggr)^2.
\end{equation}
In Table 1 we display the values of the asymmetry $\mathcal{A}_{CP}^{\prime}$ and
the numbers $N_{\st_1}$ needed to observe this asymmetry at $90\%$ CL.
  
\begin{table}[H]
\label{tab}
\caption{The $CP$ asymmetry $\mathcal{A}_{CP}^{\prime}$ defined in
Eq.(\ref{Asy})
and the number $N_{\st_1}$ of $\st_1\bar{\st}_1$ pairs
required to measure this
$CP$ asymmetry at $90\%$ CL, choosing $m_{\st_1}= 240$~GeV, 
$m_{\st_2}=800$~GeV, $m_{\SNT}=200$~GeV,
$M_2=350$~GeV and $|A_t|=1000$~GeV. For $M_{\ti Q}>M_{\ti U}$ 
we take $(|\mu|=700$~GeV, $\tan\beta=3)$ and for $M_{\ti Q}<M_{\ti U}$
we take $(|\mu|=400$~GeV, $\tan\beta=10)$.}
\begin{center}
\begin{tabular}{ccccccc}\hline\hline\\
$\mathcal{A}_{CP}^{\prime}$ & & $N_{\st_1} \times10^{-3}$ && $\varphi_{A_t}$ & & $$
\\
\\[1mm]
\hline
\\
0.05 & & 31 & & $\pi/2$ & & $M_{\ti Q}>M_{\ti U}$\\
\\
0.11 & & 7 & & $\pi/4$ & & $M_{\ti Q}>M_{\ti U}$\\
\\
0.21 & & 4 & & $\pi/8$ & & $M_{\ti Q}>M_{\ti U}$\\
\\
0.033 & & 48 & & $\pi/2$ & & $M_{\ti Q}<M_{\ti U}$\\
\\
0.030 & & 58 & & $\pi/4$ & & $M_{\ti Q}<M_{\ti U}$\\
\\
0.018 & & 165 & & $\pi/8$ & & $M_{\ti Q}<M_{\ti U}$\\
\\
\hline\hline
\end{tabular}
\end{center}
\end{table}
The cross section for $e^+e^-\to\st_1 \bar{\st}_1$ 
for $m_{\st_1}=240$~GeV is about 5$fb$ at $\sqrt{s}=$ 500~GeV
and in the range 15 -- 120$fb$ at $\sqrt{s}=$ 800~GeV, depending
on stop mixing and beam polarization.
If the branching ratio $B(\tilde t_1\to b \SNT \tau^+)$
is of the order of 5$\%$ or larger, then there are good
prospects to measure the $CP$ asymmetry $\mathcal{A}_{CP}^{\prime}$
at a linear collider with an integrated luminosity of 500$fb^{-1}$.

\section{Summary}
In this paper we have studied the $CP$ violating 
observable defined by the asymmetry of the $\tau$ 
polarization perpendicular to the decay plane
in the three--body decay $\tilde t_1\to b \SNT \tau^+$.
In the parameter domain $m_{\ti t_1}< m_{\CH^{\pm}_1}+m_b,
\ m_{\st_1}<m_W+m_{\CH^0_1}+m_b$ the branching ratio for this
decay mode can be quite large. We have calculated this
$CP$ asymmetry and the branching ratios in the MSSM 
with the parameters $\mu$ and $A_t$ complex. We give 
numerical predictions for the case of an $e^+e^-$ linear collider
with cms energy $\sqrt{s}=$ 0.5 -- 1~TeV. 
The asymmetry can reach values 
up to $\pm 30\%$. We give a theoretical estimate of the number
of produced $\st_1 \bar{\st}_1$ pairs necessary for measuring
this asymmetry at 90$\%$ CL.
There are good prospects to measure
this $CP$ asymmetry at a linear collider with an 
integrated luminosity of 500 $fb^{-1}$, if the branching ratio
of the decay $\tilde t_1\to b \SNT \tau^+$ is about 5$\%$ or larger.

\section*{Acknowledgements}

We thank M. Davier for clarifying discussions and A. Stahl for
helpful discussions and correspondence about the measurement 
of tau polarization. We also thank A.~Pilaftsis for correspondence
on 2-loop contributions to the elelctron and neutron EDMs.
This work is supported by the `Fonds zur
F\"orderung der wissenschaftlichen Forschung' (FWF) of Austria, project no.
P13139--PHY and by the European Community's Human Potential Programme
under contract HPRN--CT--2000--00149.
W.P. is supported by the Erwin Schr\"odinger fellowship Nr.~J2095 of FWF
Austria and partly by the Swiss `Nationalfonds'.

\end{document}